\documentclass[showpacs,twocolumn,floatfix,superscriptaddress]{revtex4-1}

\usepackage{graphicx}
\usepackage{bm}
\usepackage{color}
\usepackage[normalem]{ulem}
\usepackage{amsmath}
\usepackage{cleveref}

\usepackage{amssymb}
\usepackage{epsfig}
\usepackage{epstopdf}
\usepackage{gensymb}
\usepackage[utf8x]{inputenc}
\usepackage{comment}
\usepackage{ulem}

\begin{document}

\title{Two dimensional  covalent moir\'e superlattice from fluorinating twisted bilayer graphene}

\author{Depeng Ji}
\affiliation{Songshan-Lake Materials Laboratory, Dongguan, Guangdong 523808, China}
\author{Qiaoling Xu}
\affiliation{College of Physics and Electronic Engineering, Center for Computational Sciences, Sichuan Normal University, Chengdu 610068, China}
\affiliation{Tsientang Institute for Advanced Study, Zhejiang 310024, China}
\author{Lede Xian}
\email{lede.xian@mpsd.mpg.de}
\affiliation{Tsientang Institute for Advanced Study, Zhejiang 310024, China}
\affiliation{Songshan-Lake Materials Laboratory, Dongguan, Guangdong 523808, China}
\affiliation{Max  Planck  Institute  for  the  Structure  and  Dynamics  of  Matter, Luruper  Chaussee  149,  Hamburg 22761,  Germany}

\date{\today}

\begin{abstract}

Moiré materials exhibit diverse quantum properties such as superconductivity and correlated topological phases, making them ideal for studying strongly correlated systems. While moir\'e materials are typically formed by stacking two-dimensional materials with interlayer interaction dominated by weak van der Waals (vdW) forces, we explore the possibility for constructing moir\'e covalent superlattices by fluorinating twisted bilayer graphene. With first principle calculations, we demonstrate that fluorination of twisted bilayer graphene (TBG) can induce covalent bonds between adjacent layers, transforming the vdW-dominated interactions. This results in enhanced modulation of the electronic structure, with abundant flat bands across the spectrum. Our findings suggest that covalent moiré superlattices offer new platforms for exploring correlated quantum phenomena and moiré covalent chemistry.

\end{abstract}

\maketitle

Moiré materials exhibit a range of fascinating electrical, optical, and topological properties, including Mott insulators, Wigner crystals, Chern and fractional Chern insulators, superconductivity, and nematic phase \cite{cao2018correlated, cao_unconventional_2018, regan_mott_2020,li2021imaging,xu2020correlated,jin2021stripe, serlin2020intrinsic,chen2020tunable,li2021quantum,cai2023signatures,zeng2023thermodynamic,park2023observation,xu2023observation,lu2024fractional,lu_superconductors_2019,cao2021nematicity,rubio2022moire}. These features make them an ideal platform for investigating strongly correlated quantum materials \cite {balents_superconductivity_2020, devakul_magic_2021, andrei_marvels_2021,li_continuous_2021, kennes2021moire}. Moiré structures are typically formed by stacking two layers of the same two-dimensional (2D) material with a twist angle or by combining two different 2D materials with a lattice mismatch, which have been achieved in graphene, hexagonal boron nitride, metal dichalcogenides, transition metal selenides, tellurides, and so on \cite {cao_unconventional_2018, lu_superconductors_2019, yasuda2021stacking,mak2022semiconductor,lau2022reproducibility, wang_one-dimensional_2022}. The interlayer interactions in these stacked configurations are primarily governed by van der Waals (vdW) forces, which, in contrast to chemical bonds that involve electron sharing or transfer between atoms, are relatively weak. This weak interaction results in larger separations between atoms in different layers and relatively weak moiré potentials. Attempts to enhance the interlayer interaction include applying hydrostatic pressure, which shifted the superconducting transition in twisted bilayer graphene to higher twist angles and higher temperatures \cite {matthew_tuning_2019}.

\begin{figure}[h]
\centering
\includegraphics [width=0.5\textwidth] {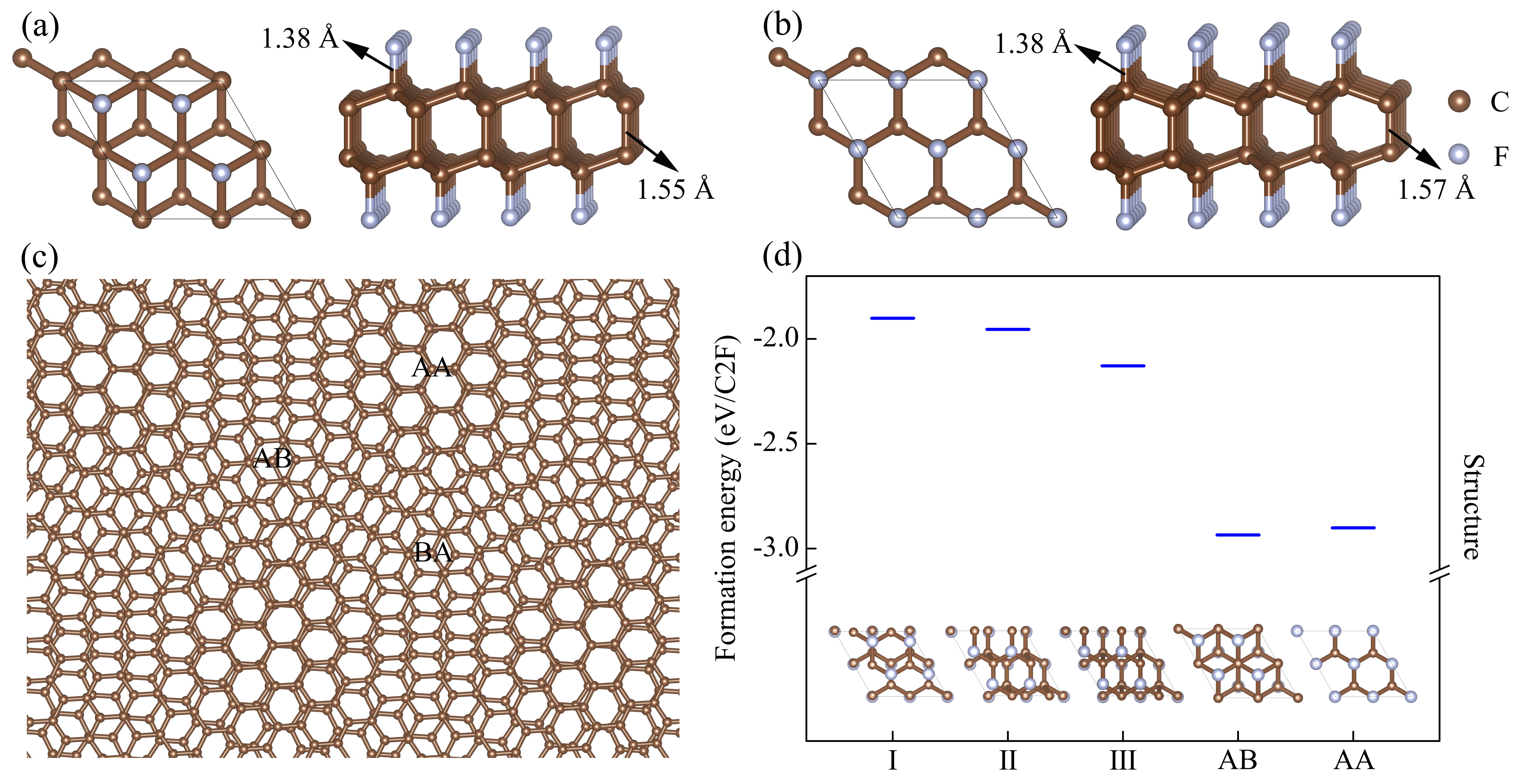}
\caption{(a-b) The atomic structures of fluorinated bilayer graphene in the AB (a) and the AA (b) stacking; (c) The atomic structure of twisted bilayer graphene showing different local stacking regions; (d) The fluorination formation energy of fluorinated bilayer graphene in different stacking geometries.}
\label{Figure 1}
\end{figure}

\begin{figure*}[ht]
\centering
\includegraphics [width=0.8\linewidth] {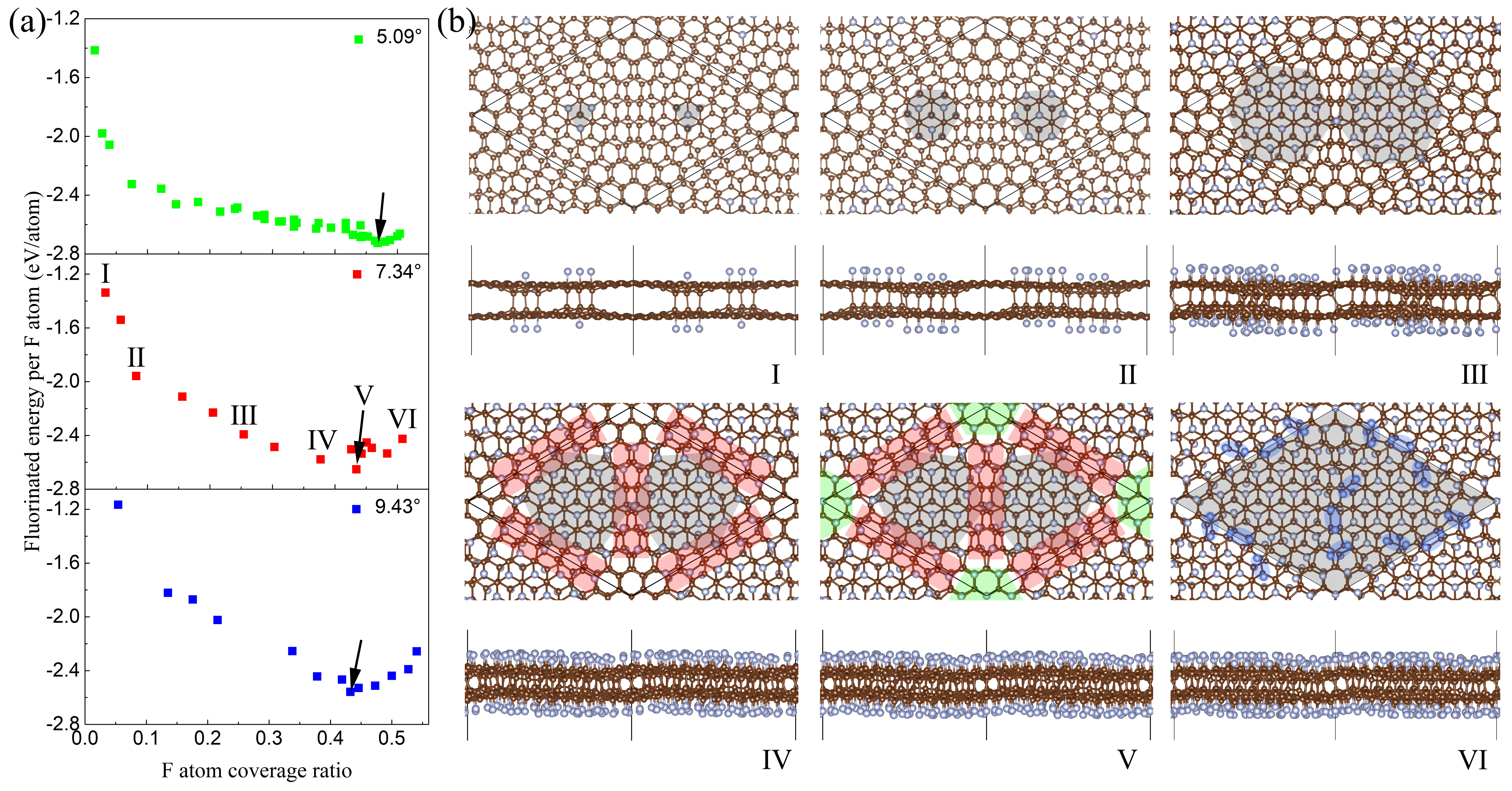}
\caption{The structural evolution of fluorinated twisted bilayer graphene. (a) the formation energy per F atom of  F-TBG at 5.09°, 7.34° and 9.43° with increasing F coverage, (b) the relaxed atomic structures of 7.34° F-TBG corresponding to the phases labeled as I-VI in (a).}
\label{Fig2}
\end{figure*}

While moiré materials dominated by van der Waals (vdW) interactions have been extensively studied, the realization of moiré superlattices with periodically modulated covalent bonds over large spatial scales remains elusive. This is primarily due to the inherent difficulty in periodically modifying covalent bonds in a controlled manner. Inspired by recent advancements in fluorinated diamane \cite {bakharev_chemically_2020,chen2022recent}, where fluorination induces the formation of chemical bonds between graphene layers, we investigate the possibility of constructing graphitic 2D covalent moiré superlattices via the fluorination of twisted bilayer graphene (TBG). Using first-principles calculations, we demonstrate that fluorination promotes the formation of chemical bonds between carbon atoms in adjacent graphene layers in TBG. The fluorination process initiates in the AB/BA stacking regions and eventually extends across the entire surface. In fluorinated twisted bilayer graphene (F-TBG), the interactions responsible for sustaining the moiré structures become covalent bonds rather than vdW forces. Compared to vdW moiré materials, this new type of moiré chemical compound exhibits significantly stronger moiré modulation of its electronic structure, resulting in an abundance of flat bands across the energy spectrum. Our findings suggest that 2D covalent moiré materials could provide a novel platform for exploring strongly correlated phenomena, moiré excitons, and moiré chemistry.

In twisted bilayer graphene (TBG), distinct local atomic stacking regions, such as AA, AB, and BA domains, exhibit varying susceptibilities to fluorination, as illustrated in Figure 1(c). Using density functional theory (DFT) calculations (details in the SI), we first evaluated the energetics of fluorinated bilayer graphene (BG) in different stacking configurations. Two representative fluorinated structures corresponding to AB and AA stackings are shown in Figures 1(a) and 1(b), respectively. Consistent with previous studies, our calculations reveal that, in the fluorinated AB-stacked bilayer graphene, the carbon atoms form a diamond-like structure with covalent bonds connecting the top and bottom graphene layers, while fluorine atoms are chemically adsorbed on the surface carbon atoms. The calculated C-F and interlayer C-C bond lengths are 1.38 Å and 1.55 Å, respectively. A similar structure is observed for AA stacking, with C-F and interlayer C-C bond lengths of 1.38 Å and 1.57 Å, respectively. Additionally, we calculated the fluorinated structures in three other stacking geometries (I, II, III) between the AA and AB stacking regions, and the results are summarized in Figure 1(d). Here we define the fluorination formation energy per F atom as: $E_{f}=(E_{tot}-E_{BG}-N \cdot E_{F})/N $.  As shown in Figure 1(d), the fluorinated structure in the AB stacking has the lowest fluorination formation energy of -2.93 eV/atom, which is 0.04 eV/atom lower than the value in the AA stacking. For the other geometries, the formation energies range from -1.9 eV/atom to -2.2 eV/atom, significantly higher than the the AB stacking value. These results suggest that the AB stacking regions in TBG are the most susceptible to fluorination.

We next examine the structural evolution of fluorinated twisted bilayer graphene (F-TBG) with increasing fluorine (F) coverage. We calculate optimized structures of F-TBG at twist angles of 5.09°, 7.34°, and 9.43°. The fluorination formation energy per F atom for these systems, as a function of F coverage, is summarized in Figure 2. A general trend is observed: as F coverage increases, the formation energy decreases until the coverage approaches 50\% (i.e., one of the two surface carbon atoms is fluorinated), after which the energy rises again as 50\% coverage is reached. Using the 7.34° system as an example, we present six representative structural phases (I to VI) in Figure 2(b) to illustrate the evolution of F-TBG as F coverage increases. Fluorination initially begins at the AB/BA regions in phase I, as these stacking configurations are energetically favorable for F absorption, as indicated by previous calculations. From phases I to III, the fluorinated area (highlighted in gray) expands until the AB/BA regions are fully fluorinated.

As shown in Figure 2(b), the chemical absorption of F atoms on the TBG surface promotes the formation of covalent bonds between carbon atoms in the two graphene layers. These interlayer covalent bonds have significantly shorter bond lengths compared to the typical van der Waals interlayer separation ($1.5\sim 1.6$ Å vs. $3.3\sim 3.7$ Å), causing notable lattice distortions. The fluorinated AB (or BA) regions experience interlayer compression, while the regions surrounding them undergo tensile strain. These strain effects extend to the AA stacking regions and the areas between AB and BA stacks. As F coverage increases and more F atoms are absorbed around the AB/BA stacking regions, the energy required to accommodate lattice distortions decreases, leading to a rapid reduction in fluorination formation energy after the initial fluorination phase.

\begin{figure}[h]
\centering
\includegraphics [width=0.5\textwidth] {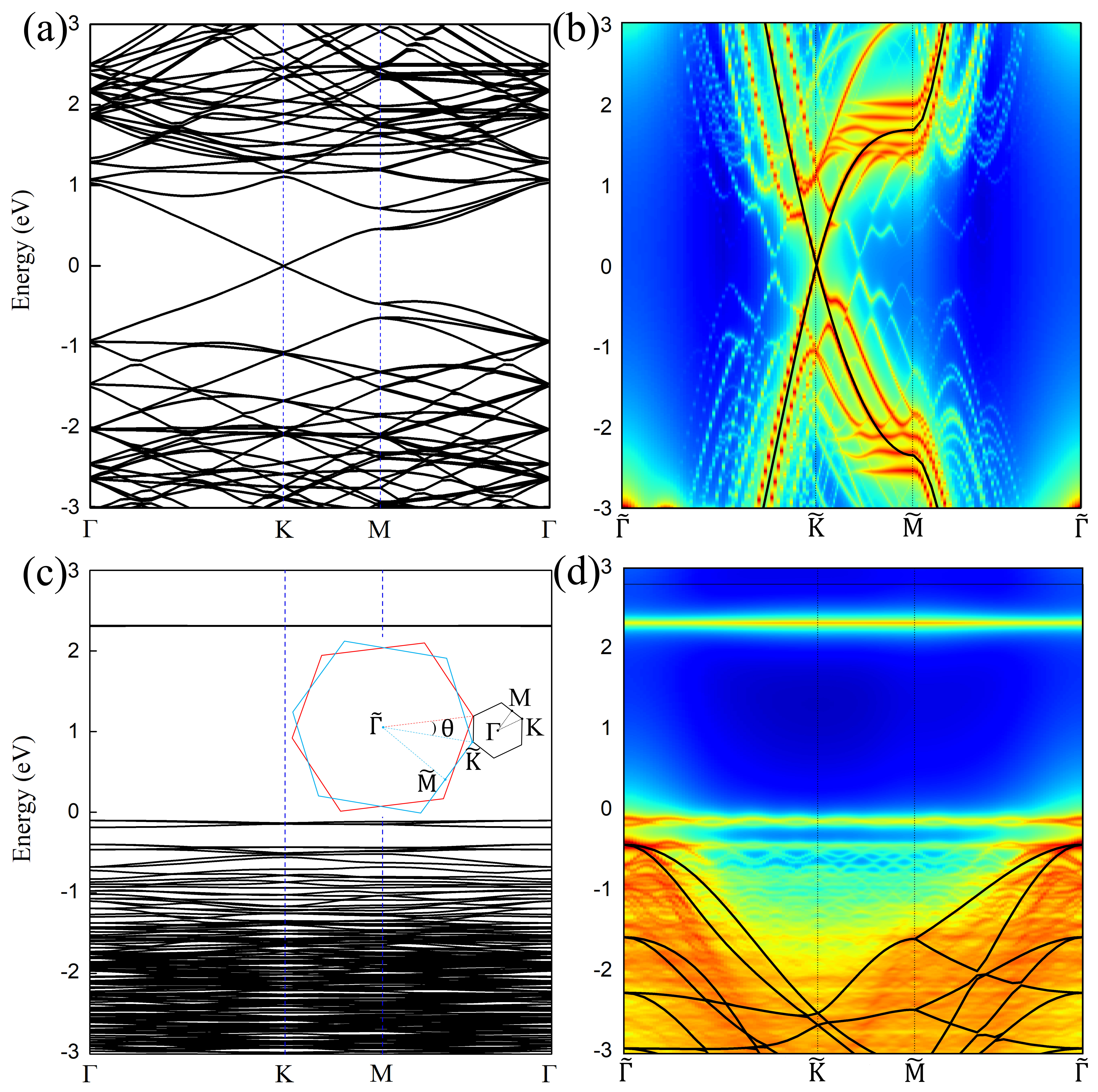}
\caption{(a) Band structure of 7.34° TBG, (b) unfolded band structure of 7.34° TBG (shown in color) in the primitive cell overlaid with the band structure of monolayer graphene (shown in black lines), (c) band structure of 7.34° F-TBG, (d) unfolded band structure of 7.34° F-TBG (shown in color) overlaid with the band structure of fluorinated bilayer graphene (F-diamane) in the AB stacking (shown in black lines). }
\label{Fig3}
\end{figure}

Once the AB/BA regions in twisted bilayer graphene (TBG) are fully fluorinated, further increases in F coverage extend fluorination into the "domain wall" regions between the AB/BA areas (depicted in red in Phase IV) and eventually into the AA regions (green in Phase V). The fluorination energy per atom reaches its minimum at Phase V, where most regions are fluorinated, but no adjacent carbon atoms within the same layer are fluorinated. As the twist angle decreases, the stoichiometric C-to-F ratio for this lowest energy phase approaches 2:1. If F coverage continues to increase, adjacent carbon atoms inevitably become fluorinated, as seen in Phase VI in Figure 2(b). This leads to additional lattice distortion and stress, lowering the chemical absorption energy of the F atoms on the surface. Post-fluorination, distinct regions of the moir\'e system exhibit a diamond-like structure (e.g., AB versus AA types), with the two carbon layers strongly bound by covalent bonds, resulting in the formation of a moir\'e covalent compound. 

As the weak van der Waals interactions between layers are transformed into strong covalent bonds in fluorinated twisted bilayer graphene, its electronic properties are expected to change significantly compared to those of the original twisted bilayer graphene. The band structures of 7.34° TBG and the most energetic favorable phase of F-TBG (Phase V in Figure 2) are shown in Figure 3. The band structure of TBG exhibits semi-metallic characteristics, with a linear crossing (the Dirac cone) located at the Fermi level,  similar to that of monolayer graphene. However, in F-TBG, the Dirac crossing disappears, and a relatively large band gap emerges instead. Additionally, F-TBG exhibits abundant flat bands throughout its entire band structure. This is in contrast to the relatively dispersive band dispersions in TBG at the same twist angle.

To further analyze these changes, we perform band unfolding of both TBG and F-TBG into their untwisted primitive Brillouin zones and the results are shown in Figures 3(c) and 3(d). The unfolded band structure of TBG closely resembles that of monolayer graphene (indicated by the black solid lines in Figure 3(c)), whereas the unfolded band structure of F-TBG is more similar to that of fluorinated diamane (Figure 3(d)). These results suggest that after fluorination, the C-C \emph{sp}$^2$ hybridization of graphene transforms into \emph{sp}$^3$ hybridization in F-TBG, similar to F-diamane \cite{bakharev_chemically_2020,lavini2022two,boris2021}. Moreover, it can be seen from these unfolded band structures that the band dispersions of TBG retain most of the monolayer characteristics in the primitive cell and only open gaps where replica bands intersect. Thus, the electronic properties of TBG can be described using monolayer Hamiltonians of the top and the bottom layers coupled with interlayer moir\'e potentials terms. The low-energy states of TBG (at twist angles larger than the magic angle) can be treated perturbatively, manifesting as Dirac electrons with a renormalized Fermi velocity \cite{tbg2007,macdonald2011}.  In contrast, the unfolding band structure of F-TBG is very blurry, indicating that the electronic states are strongly affected by the moir\'e potential, rendering the perturbative approach invalid for F-TBG across the entire band structure even at relatively large twist angles. The difference in the moir\'e potential strength between TBG and F-TBG can be attributed to their distinct origins: in TBG, the moir\'e potentials are primarily influenced by weak van der Waals interactions and strain from atomic reconstruction, while in F-TBG, they are dominated by variations in strong covalent bonds at different domains.

\begin{figure*}[ht]
\centering
\includegraphics [width=0.95\textwidth] {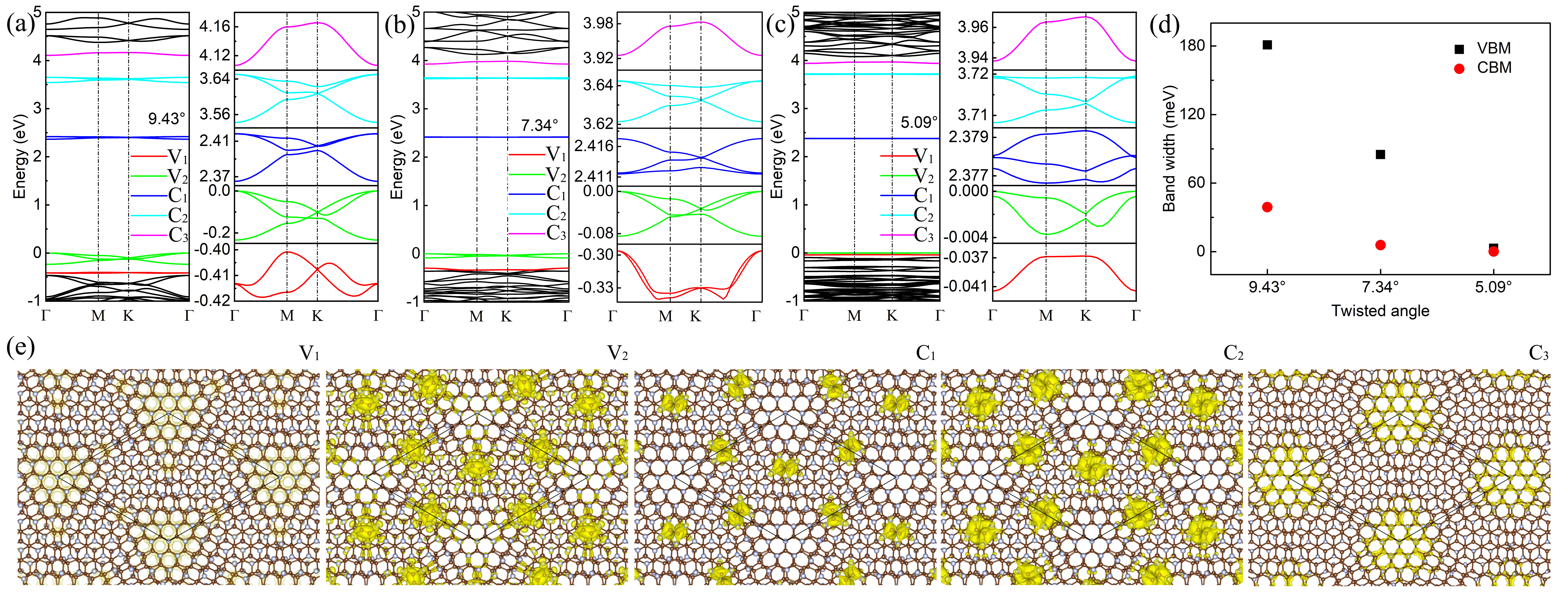}
\caption{Band structure evolution of fluorinated twisted bilayer graphene at different twist angles. (a-c)Band structures of F-TBG at 5.09$^{\circ}$, 7.34$^{\circ}$ and 9.43$^{\circ}$ with the five sets of flat bands around the Fermi level highlighted with color, (d) evolution of the bandwidth of flat bands at the VBM and the CBM with decreasing twist angles, (e) charge density distribution of the five sets of flat bands in F-TBG at 7.34$^{\circ}$.}
\label{Fig4}
\end{figure*}

We further analyze the twist-angle dependence of the electronic states in F-TBG at their most energetic favorable phases. The calculated band structures of F-TBG at twist angles of 5.09°, 7.34° and 9.43° are presented in Figure 4. In particular, we focus on the five sets of flat bands at the band edges highlighted with color in Figures 4(a-c), which include top two sets of valence bands (V$_1$, V$_2$) and bottom three sets of conduction bands (C$_1$, C$_2$, C$_3$). A general trend for the flat bands in F-TBG is that the bandwidth decreases as the twist angle decreases. For example, the bandwidth W of V$_2$ at the valence band maximum (C$_1$ at the conduction band minimum) decreases from 181 meV (39 meV) at 9.43$^{\circ}$ to 4 meV (2 meV) at 5.09$^{\circ}$ as shown in Figure 4(d). These bands are significantly flatter than those in TBG at the same twist angles. At a twist angle of 5.09$^{\circ}$, the flat bands at the band edges of F-TBG already exhibit a bandwidth W of less than 4 meV, even flatter than the bands observed in TBG at the so-called magic angle. The onsite Coulomb interaction U can be estimated as $\frac{e^2}{4\pi \epsilon d}$, where $\epsilon$ is the  the effective dielectric constant, and $d$ is the effective linear dimension of each moir\'e site that is proportional to the moir\'e supercell size and inversely proportional to the twist angle. Consequently, U is larger for systems at larger twist angles. It can be expected that the correlation effects in these covalent moir\'e systems at large twist angles are significantly stronger than those in TBG at the magic angle due to the larger U and smaller W. 

The flat bands in fluorinated twisted bilayer graphene (F-TBG) exhibit intriguing electronic properties, with dispersions that can be modeled using simple prototype lattice frameworks. For instance, the C$_3$ and C$_2$ bands of F-TBG at a twist angle of 7.34$^{\circ}$ display characteristic dispersions that correspond to the single-orbital triangular lattice and single-orbital kagome lattice models, respectively, while the V$_1$ band dispersion aligns with the px-py two-orbital triangular lattice model (see figure 4(b)). These dispersion characteristics are corroborated by calculations of charge density distributions shown in figure 4(e), which reveal triangular and kagome lattice patterns for bands C$_3$/V$_1$ and C$_2$, respectively. Consequently, these flat bands in F-TBG provide a platform for simulating exotic correlated physics related to triangular and kagome lattice models. Notably, certain flat bands exhibit electronic states localized at the domain wall regions between AB and BA domains, with band energies residing within the bulk band gap of fluorinated diamane. In particular, the C$_1$ bands appear as deep "defect" states, with an energy separation exceeding 1 eV from the conduction bands and over 2 eV from the valence bands. These flat band states may function as color centers similar to those in diamond and other semiconductors, presenting potential applications in quantum sensing, single-photon emission, and quantum computing \cite {childress_coherent_2006,nvcenter2014,hepp2014,pezzagna2021,Castelletto_2020}.

In conclusion, we propose a novel class of moir\'e materials, termed moir\'e covalent superlattices, which can be synthesized by functionalizing moir\'e van der Waals materials. We demonstrate the feasibility of this approach by investigating the functionalization of twisted bilayer graphene (TBG) with fluorine using density functional theory (DFT) calculations. Our findings indicate that the weak van der Waals interactions between graphene layers in TBG are transformed into strong covalent bonds upon fluorination. As the coupling strength between layers becomes comparable to that within layers in fluorinated TBG, moir\'e superlattice effects significantly alter the properties of all valence electronic states in the system, resulting in the formation of flat bands across the entire energy range, even at relatively large twist angles. The emergence of ultraflat bands at the band edges and within the band gap positions these new moir\'e materials as excellent platforms for exploring strongly correlated physics and for potential applications in quantum sensing, single-photon emission, and quantum computing. Our proposal of constructing moir\'e covalent superlattices by functionalizing moir\'e van der Waals materials could also apply to twisted multilayer graphene and other twisted 2D materials \cite{lavini2022two}, opening new pathway for the study of moir\'e covalent chemistry.

\section{\textbf{ACKNOWLEDGEMENT}}
This work is supported by the National Key Research and Development Program of China (Grant No. 2022YFA1403501), Guangdong Basic and Applied Basic Research Foundation (Grant No. 2022B1515120020), the Hefei National Research Center for Physical Sciences at the Microscale (KF2021003), the National Natural Science Foundation of China (Grant No. 62341404), Hangzhou Tsientang Education Foundation and the Max Planck Partner group programme. L.X. acknowledges the support by the startup grant of Songshan Lake Materials Laboratory. The calculations were performed using resources provided by the Max Plank Computing and Data Facility, as well as the Platform for Data-Driven Computational Materials Discovery of the Songshan Lake Materials Laboratory. 

\bibliography{ref}

\end{document}